# Managing Situations of Complexity and Uncertainty

# The Contribution of Research and Development

L. E. Brunet[1][2], E. Longcôté [3]

[1] R&D Mediation, 71 av. de Lattre de Tassigny, Bourges, France

[2] Associate researcher at the University of Clermont Auvergne

[3] APEKSA Consulting

**Abstract**

*The second industrial revolution saw the development of management methods tailored to the challenges of the times: firstly, the need for mass production, and then, the pursuit of improved quality and customer satisfaction, followed by a push to improve operational performances in response to market globalization. If these approaches were initially inspired by rational mechanistic thinking, they have since gradually broadened to integrate other dimensions such as psychology, sociology and systemic analysis. Business enterprises underwent a profound rethink in the 1990s introducing increasingly refined modi operandi, only to find their environment disrupted by the appearance of two new parameters: complexity and uncertainty. Enterprises of the third industrial revolution were able to integrate these parameters at the outset, introducing new styles of management. However, these may well be deficient with regard to activities where an error may be fatal, or a failure intolerable. Caught between the precautionary principle and the principle of experimentation, the third industrial revolution falters to find the right approach, whereas the fourth industrial revolution is almost already upon us, bringing its lot of upheavals. In this regard, faced with increasing complexities and uncertainties, Research and Development is of particular interest since its vocation consists precisely in confronting the complex and the uncertain. This article examines the fundamental principles of the R&D process, and analyses how these may act as a benchmark for contemporary management by providing sources of inspiration.*

**CONTENTS**

**BIBLIOGRAPHY**

# 1 A HISTORICAL STUDY OF MANAGEMENT

Without claiming to be exhaustive, here follows a historical summary of the main trains of thought behind modern management approaches from the beginning of the 20$^{th}$ century, and of its emblematic figures.

## The Classical Period

At the turn of the 20th century, Frederick Taylor was the first to theorize principles of organization, within a scope limited to production workshops [1]. He argued that in addition to the classical vertical hierarchy, a horizontal division of labour, breaking up and specialising each task, would enable each to be performed with optimal efficiency. The "assembly line" appeared, which Charlie Chaplin will display twenty-five years later in its film "Modern Times".

Contemporaneously, in France, Henri Fayol developed a managerial approach that encompassed the enterprise in its entirety. This comprehensive view of management became known as "administrative management" [2]. In Germany, Max Weber defined an organisational approach later termed "bureaucratic management" [3].

All three approaches embody the "command and control" model inherited from monarchical institutions, where the military aristocracy and clergy commanded, leaving the common people to supply the workforce and obey.

## Human Relations

The 1920s saw the advent of a new era: that of Human Relations. Mary Parker Follett was one of its prominent pioneers. Drawing on her long experience in social work in Boston, she widened the scope of management at a single stroke [4]. For the first time, the human factor was included, which allowed radically disruptive notions to emerge: the wealth of diversity, the advantage of constructive conflict, the virtue of "win-win" negotiation, the limitations of salary as a motivating factor and the belief in other levers: participation, empowerment, autonomy.

A little later, Elton Mayo and his famous Hawthorne studies demonstrated that motivation was influenced more by subjective factors than objective ones, namely those of consideration, recognition, and identification with a group.

In their wake, Abraham Maslow later developed his theories into what would become widely known as Maslow's hierarchy. In the 1960s, Douglas McGregor theorised two motivation streams [5], distinguishing classical so-called "extrinsic" motivation of the carrot-stick type, from so-called "intrinsic" motivation which stems from the nature of the work itself.

Thanks to these thinkers, business companies underwent a change of dimension: psychological needs were gradually being taken into account, motivation steadily gained ascendance as a key factor in management.

## Systemic management

As the Second World War came to its conclusion, the management body of knowledge had already expanded considerably, but the majority of western business and industrial companies were still applying a Taylorist approach to their activities. Essentially, focus remained firmly fixed on how the enterprise worked on an internal point of view, with little regard to the outside world. However, management was on the cusp of a major rethink: the environment became a factor to be borne in mind. The enterprise was increasingly perceived not as a closed entity but as an open *system*.

The major figurehead of this so-called systemic current of thought was Peter Drucker [6]. He argued the importance of the stakeholder, in particular the customer, as well as two other key functions of the company: marketing and innovation. Following in the footsteps of Mary Parker Follett, he invented the concept of participative management by objectives (PMBO).

Another prominent figure of the times was William Deming, known as one of the forefathers of Quality. Often used simplistically, "Quality" should be understood in its wider sense: the capacity to manage an enterprise efficiently and to satisfy its customers, beyond merely ensuring the quality of its products. Deming's work gave birth to notions that have since become immensely widespread: operational performance (which today tends to be termed "operational excellence"), management by process, continual improvement (Plan-Do-Check-Act or PDCA), and in all cases defended the precedence of management over technical skill.

From their combined works, two notions arose that would affect society as a whole: the systemic approach, premise to sustainable development and Corporate Social Responsibility (CSR), and the knowledge-based economy that would be revolutionized by the explosion of the Internet and the New Technologies of Information and Communication (NTIC), whose impact would be felt from the 1990s onwards. Embracing the individual in relation to skills development and its contribution to processes, these concepts had little regard for organizational structures or human relations. A fourth current of thought that developed during the 1960s-1970s will remedy this: the sociology of organisations.

## The sociology of organisations

Michel Crozier, in France, was one of the pioneers of this movement. His studies plunge us into the depths of human functioning ([7] and [8]) and the enterprise changes its face to become "the realm of power relations, influence, bargaining and calculation" [8]. Faced with this more sombre reality, he considered that the role of the manager is not to comply with it, nor to suffer it, but to help it evolve upwardly. His work introduces a keystone of contemporary management: the management of change.

In the United States, sometime before, James March was working on similar concepts [9]. Apart from coming to the same conclusions regarding power games, decision-making and the central importance of negotiation, he also underscored the importance of learning lessons from what the organization actually produced (its choices, its successes, its errors), which he termed "organizational learning". The latter shows the emergence of feedback not merely from a process perspective, as advocated by William Deming to feed the continual improvement system, but from the perspective of the organization itself, viewed as a living organism.

This current of thought would later be enriched by the works of two great French thinkers: Jean-Christian Fauvet, founding-father of socio-dynamics [10], and Edgar Morin.

## Strategic management

By this time, the scope of management had been considerably enlarged, however, one crucial dimension had not yet been taken into account: the market, the strategy.

Igor Ansoff, with his eulogy to strategic planning, which was hugely successful, performed the first studies in this direction in the 1960s. Unfortunately, the ever faster evolution of the business environment from the 1970s, at an ever increasing scale, made this approach largely utopic, being intellectually appealing but by nature static despite periodical revisions. Planning was thus abandoned in favour of a more flexible vision, both versatile and interactive: strategic management.

Henry Mintzberg was the first great proponent of strategic management [11]. He defined the so-called 5Ps of strategy (Plan, Pattern, Ploy, Position, Perspective), which integrated the SWOT approach. He advocated the participative development of strategy, enriching the top-down "deliberate" vision by a bottom-up "emergent" vision from the line managers. Last but not least, the customer was once again the central focus, and the eminently contingent nature of management was underscored.

A second prominent figure of this current of thought, Michael Porter, defined another tool for strategic development dubbed "Porter's five forces". In his approach, strategy consists in the constant search for competitive advantages, and their preservation. To this end, innovation plays a core role and acts as a major lever in distancing or discouraging competitors, by setting up "entry barriers". He also created a model of value chain analysis, assessing the contribution of the enterprise's different functions to its operating profits.

## Recent currents of thought

In the early 1990s, the concept of management pursued its evolution in close symbiosis with society as a whole.

From a societal perspective, the notions of CSR and sustainable development were in full swing, despite the frequent contradictions between displays of political correctness and practical realities. The practice of risk management was increasingly prevalent, in response to the precautionary principle, and would permeate to varying degrees according to the liberality of each State and the expectations of public opinion, a point to which we will return later.

From a strategic perspective, the last three decades have been marked by the globalization of different approaches and reflection on the enterprise's core business, leading certain parts of their activity to be relocated or outsourced, as well as by the increasing pressure of short-term rationale to the detriment of the long-term.

From an operational perspective, these same decades have been marked by a tightening up of performance management, which has become central to the functioning of the enterprise. This is apparent from the rise in standards and normative guidelines of the ISO type, the adoption of management systems, the deployment of information systems, the multiplication of improvement approaches such as Lean management and process re-engineering, the emergence of the notion of agility, firstly within the scope of software projects and then more globally, the principles of which gave birth to a manifesto in 2001.

From a human perspective, this period has seen the development of the HR function, which has expanded its role beyond the mere management of personnel to encompass vocational training, the diversification of motivational levers (performance reviews, profit-sharing, incentive schemes, employee savings plans, employee benefits, etc.), the development of transversal multicultural management, company restructuring and globalization.

All this has undeniably generated a gain in efficiency, but has also led to adverse effects: a divorce between the centres of decision-making and the operational actors at working level, an inflation in procedures, constraints, controls, reporting, a loss of sense and motivation, and the generalization of stress and burn-out. In management parlance, engagement (or empowerment) has replaced motivation as the new leitmotiv. Numerous studies, and in particular that of the GALLUP Institute [12], have revealed a disturbing tendency : in 2013, only 14% of European employees considered themselves "engaged", compared with 66% "disengaged" and 20% who consider themselves "actively disengaged" (which is to say, dissatisfied, unhappy and likely to spread their negativity to their co-workers). In France, these findings are even more alarming.

In response to these warning signs, new currents of thought have appeared, advocating not just to adjust management styles but rethink their paradigms totally. The quality of working life (QWL) movement, as well as the so-called liberation management or the "opal" managerial philosophy, both embody such trends.

Returning to the basic psychological needs which traditional companies no longer fulfil, these new currents emphasize employees' autonomy and participation, the development of their skills and their potential, the loosening of functional constraints, and the importance of collective intelligence. QWL focuses on both collective and individual well-being, and may bring out the notion of happiness at work, leading to new functions being created such as that of Chief Happiness Officer. Liberation management or the "opal" management, respectively endorsed in France by Isaac Getz [13] and Frédéric Laloux [14], promote new managerial paradigms: trust, self-management, collegial decision-making and, borrowing from the works of Robert Greenleaf [15], redefine leadership according to the notion of "servant-leader". Occasionally accused of being naively optimistic, these currents do not advocate the creation of well-being or liberation as an end in itself but as keys to empowerment, efficiency, and innovation. Certain authors also underline the potential drifts that may affect such organizations, despite the inherently virtuous nature of their paradigms [16].

At the risk of over-simplifying by reducing management to two dimensions, the process vs the human on the one hand, and the internal vs the external on the other, the following diagram illustrates the main stages in managerial thinking (Figure 1):

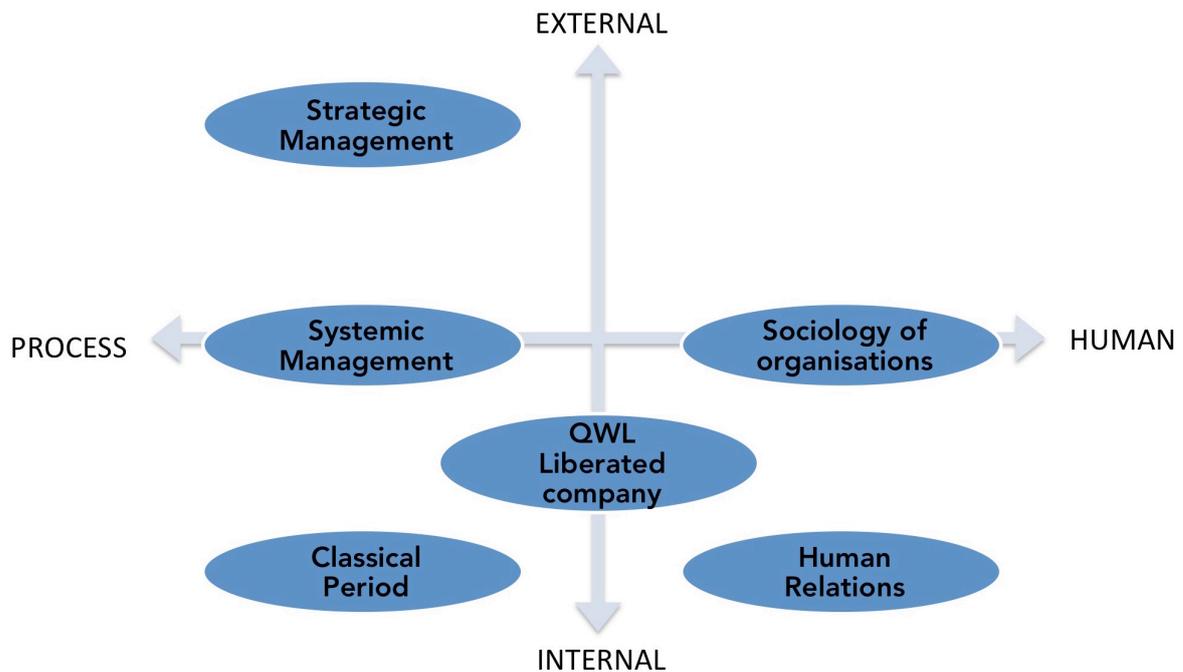

**Figure 1: Panorama in managerial thinking**

## 2   CONTEMPORARY ENVIRONMENT

After this retrospective study, let us examine the present situation and then cast our minds to the future. We will firstly attempt to determine the main lines of the incipient industrial revolution, positioning it with respect to the three revolutions having preceded it, and then review the major characteristics of our contemporary environment. Finally, we will examine the issues that are emerging in terms of management.

### 2.1   Fourth industrial revolution or revolution 3.1

The term "industry 4.0" was coined by the founder of the Davos economic forum, Klaus Schwab in his book entitled "the Fourth Industrial Revolution" [17]. Let us rapidly run back over the first three, given that an industrial revolution is defined by the concomitant appearance of a new energy and a new communication technology. During the first two revolutions, these inter-linked elements (energy, communication technology) were respectively the steam engine and printing presses, then electricity and tele-communication means (telephone, telegraph, television). The third industrial revolution, which made its appearance in the early 1990s, was that of network functioning and web communication coupled with an energy distribution system approaching that of http protocols, namely smart grids (Table 1).

**Table 1: The industrial revolutions**

|   |   | ENERGY | COMMUNICATION | PRODUCTION SCHEME | OPERATOR |
|---|---|--------|---------------|-------------------|----------|
| 1ST | 18th-19th | Steam machine (Watt) | Machine-driven rotary presses – mass-circulation newspapers | Linear | Scalar |
| 2nd | 1871-1914 | Electricity (Edison) | Telephone – telegraph – television | Tree structure (System) | Matrix |
| 3rd | 1990 | Power grids- Smart grids- Hydrogen- Nuclear Fusion perhaps soon | Web (Tim Berners-Lee)- Internet- Social networks- P2P- AI | Graphs and networks | Tensor |

According to Schwab, the advent of a fourth revolution is proclaimed by the arrival of nano-bio-info-cogno technologies, artificial intelligence (AI) or the use of robots. Although the term fourth revolution is commonly employed, L. Ferry has established that it is being misused here [18]: indeed, there is nothing revolutionary from an energy point-of-view other than the reinforcement of the ubiquitous power-grid introduced by the third industrial revolution. However, what we experience today is truly original, akin to a renaissance, and goes beyond a mere iteration of industrial practice. Indeed, contrary to what happened during the first two industrial revolutions and as was the case for the Italian Renaissance period, it is more a question of intelligence product – the printed book, erstwhile, and the web today – rather than energy. The current effervescence regarding compact fusion reactors, of a size no greater than that of a truck but able to supply enough power for a city of 100,000 inhabitants, could revolutionise energy production bringing us de facto into the fourth revolution within the very short term (5 to 10 years).

Hereafter, the term « fourth industrial revolution » will be used even if strictly speaking it is not yet effective.

## 2.2 The VUCA world

One current of thought distinguishes four key characteristics of our contemporary environment [19]:

- Volatility, which is to say the unpredictability of situations encountered, their likelihood for rapid evolution at greatly variable magnitudes,

- Uncertainty, which is to say the impossibility to understand the situation objectively, to establish cause to effect relationships, and foresee the repercussions of decisions,
- Complexity, which is to say the plethora of parameters influencing the situation, their interdependence, their interaction
- Ambiguity, which is to say the impossibility of having a single interpretation of the situation.

This is the "VUCA world".

This concept was introduced into the American Army after the end of the Cold War, when traditional paradigms had disintegrated. Later, it would be rekindled from the ashes of the conflicts in Afghanistan, Pakistan and Irak, where drastic changes in the theatres of operation had to be confronted. In his blog [20], Philippe Silberzahn rightly points out that this acronym conflates terms of contrasting natures: volatility and complexity are characteristics that stem from our environment, whereas uncertainty and ambiguity lie within the limitations of human cognition. It is important, therefore, to distinguish between what is within the scope of human dependency, and what is not.

Post 1968 social movements, the oil crisis, galloping globalization, the collapse of the communistic block, the emergence of BRICS on the international scene, the revival of religious fundamentalism, the bursting of speculative bubbles, the subprime crisis, climate disruption, the expansion of the web, NTICs, genetic engineering and AI, all of the above illustrate, over a remarkably short period of time, the end of a world formed of blocks deemed to be stable, and the beginning of this VUCA world.

In the remains of this article, we will focus on two of its components: complexity and uncertainty.

## 2.3 New issues in management

The notions of complexity and uncertainty are not the prerogatives of the third industrial revolution or of the VUCA world. As discussed above in the retrospective analysis, the business enterprise was subject to considerable evolutions of increasing complexity and uncertainty throughout the 20th century.

From an internal perspective, in 1950s, the sociology of organisations revealed a new reality of the enterprise that had been largely overlooked up to that point: it was no longer a coherent and stable entity but rather the seat of tensions, perpetual migrations, where management becomes in essence *contingent*. The role of directors and managers evolved into something more complex, more challenging, or even almost schizophrenic: it entailed managing the tensions between explicit authority and nebulosity, the contradictions between the objectives of the organization and personal objectives, and trying to keep an even keel at the same time as pushing through changes. To succeed in this, directors and managers could no longer claim to hold the truth, nor to convince once and for all, nor to forcibly assert themselves, but could only rely on negotiating, confronting qualms, inertia, resistance, and creating new flows. In other words, they had to abandon their erstwhile "governing powers" and reinvent new power modes.

Thereafter, the apostles of strategic management found little interest in the operational and relational aspects of management, but were fully aware of the difficult role of the manager. As Henry Mintzberg argues: "managers do all the dirty work: they sort out the difficult problems and the complex connections. This is why the practice of management is so ambiguous and why words such as experience, intuition, judgment and wisdom are often required to describe it. Gather together a good quantity of craftsmanship, add a touch of art, sprinkle it all with science and you will obtain a result that is above all a practice. And remember: no management method is better than any other. It all depends on the situation." Ultimately, managers are left both more strategist and more alone.

In the 1980s, the notions of complexity and uncertainty went up a notch: the volatility of shareholders and directors, the diversity of organizational options (partnerships, industrial restructuring, matrix or project organisations, etc.), the increasing multidisciplinarity of projects, the pressure of working in the short-term and improving performance, the never-ending optimization of company processes, created an ever-changing internal landscape where it seemed no longer possible for anything to be conclusively controlled, or acquired.

From an external perspective, the emergence of systemic management in the 1950s represents a first major step in complexity and uncertainty: the enterprise was no longer considered as a closed entity but an open system, exposed to a multitude of influences and vagaries over which it had little or no control. The inception of strategic management was a second step: the company was deemed to take account of a market that existed in and of itself (reaction) and to try and exert an influence on said market (pro-action), akin to the squaring of the circle.

In short, complexity and uncertainty are not new factors for an enterprise today; however, the fourth industrial revolution and the VUCA world have caused them to develop in scale, intensity and impact.

Within this scope, one last point must be mentioned: the human feels threatened by the non-human. There again, it is not a new debate. Man and machine have enjoyed conflicting but passionate relations since the 19th century. However, even as the fourth industrial revolution amplifies these phenomena, it also lends an unprecedented depth to this issue: humans are no longer merely competing with machines or robots, but also with so-called artificial creatures claiming what humans believe they alone enjoy and what differentiates them from the rest of Creation: intelligence.

On the heels of a century of evolution, this new reality calls once again for managerial practices to be reconsidered, but this time much more radically.

# 3 MANAGEMENT UNDER REVIEW

We will start by highlighting the limitations to previous managerial models with respect to the aforementioned stakes. After this, we will examine the attitude to be taken: should we "tweak" existing practices once more, or should we adopt new paradigms?

## 3.1 The limitations to previous models

The second industrial revolution was founded on the mechanistic approach, namely two general principles of a Cartesian nature: the relation of causality (a cause produces an effect), and the possibility of breaking a system down into parts. As we have seen, this mechanistic approach has already shown its limitations and, since the mid-1950s, the enterprise has been considered not as a machine but as a system, a living organism: the causes do not produce their effects alone, the effects have retro-effects on the causes, systems have attractors – in the mathematical sense of the term – and have emergent properties, which is to say that the sum of their parts constitute a whole having new properties [21]. And to further complicate matters, components proceeding neither from reason, nor analysis, nor determinism, have forced their way: emotions, motivations, aspirations, beliefs, relationships and co-existence. In parallel, the world outside the enterprise started to evolve at an unprecedented rate during the 1970s. Faced with this impetus for change, both internal and external, the enterprise adapted its modus operandi.

From a strategic perspective, as seen above, planning was quickly abandoned in favour of a more flexible and adaptable approach: strategic management. If we look more closely, we see that if the principle of long-term forward planning has been largely abandoned (at least in the private sector, the public sector still retaining this practice), strategy has continued to be supported by that king of all tools, also based on forward planning: the business plan. Essential for communications, indispensable for banks when deciding to support this or that project, it has become the alpha and omega of strategic management. Apart from that, the same tools are used as before, opinion polls, marketing studies, extrapolation-based models, wherein the future can be extrapolated from a study of the past and present. That these tools have not been able to prevent errors, certain of which are well known [22], is a clear indication of their inadequacy. Lately, the subprime crisis was a dramatic illustration of the collusion between lies and models, the latter being ill adapted but even worse, misunderstood by governing bodies.

From an operational perspective, several disciplines have become the new queens of management since the 1960s.

Such is the case for risk and safety management, within a societal context marked by the increasing importance placed on human life, the pressure of total safety and the precautionary principle. This consists in mapping incidents and accidents likely to affect the enterprise, its activities, its employees, its projects, its investments, in characterizing them by their likelihood, severity and detectability index [23], and in formulating remediation strategies. There again, previous methods continue to be used: modelling, anticipation, extrapolation, leading to an inflation of "plans" of all types (prevention plans, mitigation plans, business continuity plans, corrective action plans, etc.). Mobilizing generations of engineers, this discipline has been costly but has great appeal: technicians saw in it a rehabilitation of reasoning, managers an aid in decision-making, and society a way to exemplify its values, and overcome its fears. Effective in dealing with known hazards, it is, however, ineffectual with regard to disruptive contexts, to emergent risks that have not been previously encountered. Intellectually-appealing, reassuring, but unable to take into account the realities of today's world, it is largely a sham, an illusion: most systems are complex, in the thermodynamic sense of the term [24], and just do not let themselves be addressed through such methods.

Another discipline mentioned previously has also been extremely relied on: process re-engineering. This consists in modifying certain processes to curb any anomalies detected, improve operational performances, or pursue new objectives. There again, the approach is based on previous paradigms: repeatability (the essence of a process lies in the repeatability of its elements), planning, reinforced by additional action plans regarding the organization, the tasks, the skills, the tools, the information system, etc. Equally virtuous in advancing both operating modes and collective culture, in mobilizing all actors around common goals, in breaking down barriers, rallying, empowering, it forces the enterprise to make a crucial decision: either merely adapt to the existing situation (logic of continuity), or else totally rethink the situation (logic of disruption). In practice, everything tends towards the first option, which seems less costly, less time-consuming, and less risky. Therein lie the limitations of this approach. Moreover, everything tends do deter it: the rules, procedures, controls, dashboards, multiplicitous plans take over, sacrificing the human on the altar of performance and profitability.

To sum up, the movements made on a strategic and operational level both force methods to evolve, whilst largely retaining previous representations. They also demonstrated a tendency to follow fashionable trends: appearing in successive waves, so-called "solutions" have ensnared both decision-makers and managers, over-eager to adopt them without necessarily making the effort to understand the ins and outs, happy to apply them in a copy-paste way without worrying about the particular context or identity of their enterprise.

The enterprise of the second industrial revolution certainly attempted to adapt its modus operandi, and didn't spare its efforts. However, this was often done without sufficient in-depth analysis, privileging technical processes to human and cultural dimensions. Certain currents of thought mentioned in the background discussion above, namely quality of work life, the liberated enterprise or "opal" enterprise, took that extra step, questioning inherited paradigms. Having said that, these recent ideas are not mainstream.

## 3.2 Evolution of the paradigms

In the 1990s, the third industrial revolution grew out of the boom in information technology and was marked by the constitution of very complex systems, very emergent, revealing even further the limitations of prior approaches. Studied as early as 1920 [25], the relation between profit and the capacity of organisations to manage uncertainty were particularly highlighted. Thus, none of the GAFA (Google, Apple, Facebook, Amazon) are built upon business plans, contrary to what is currently taught in schools of management, but on largely speculative and exploratory business models, co-constructed with the stakeholders and constantly confronted with market realities. A. Osterwalder thus proposed a new approach organised along nine themes [26] to replace classical business plans. Beyond that, the whole methodological reference system inherited from the second industrial revolution underwent a total rethink.

If risk and safety management remains operative in traditional industrial enterprises, digital enterprises are more interested in opportunities rather than risks. Building on the success of Paypal, P. Thiel [27] made uncertainty the new key to success, provided that it is managed correctly and there are no regrets for the "world of yesteryear".

Similarly, if process optimization remained a stable of established enterprises, a new concept swept through emerging companies: innovation. Naturally, innovation existed before. On a managerial plane, innovation was present in every stage of our retrospective analysis. Technologically speaking, the 30-year post-war boom period was rich in innovation. Strategically, Michael Porter considered it a major strategic lever: "innovation is the key to economic prosperity". To that end, he defended the notion of "coopetition", an ad-hoc alliance between former competitors in certain domains or on certain projects, as well as the network operation between universities, research laboratories and funding bodies, that became generalized in the United States in the form of business clusters before reaching France in the form of competitiveness clusters. In short, innovation is in no way the prerogative of the third industrial revolution, but the latter has considerably reinforced its stakes and its scope: it is no longer a question of improving what already exists, of innovating products or services within the traditional scope of technology, but of innovating throughout the whole enterprise, the business model, the marketing, the processes, the social aspects.

The methods and tools, the former queens of creativity and collective intelligence, the most fashionable of which being brainstorming, have been scientifically proved to be ineffective [28][29][30] at best, or even counter-productive as explained by the Dunning-Kruger [31] cognitive bias effect at their worst. These methods therefore tend to be replaced by design thinking [32] methods, wherein the first of five forces involves guided interviews of the different stakeholders, using Porter's definition of this term [33]. Lastly, dashboards and performance indicators have fallen from grace with managers, the idea no longer being to demonstrate growth or improvement with respect to a known reference, but to create new flows entirely.

Thus, far from being a simple iteration, the third industrial revolution stems from a Copernican revolution: the unknown and the complex become established, past solutions are no longer effective, traditional benchmarks waver as certainties fail. In many ways, the process follows that of grief: we must mourn the loss of our stability, security, tried-and-tested intellectual approaches, comfortable, reassuring but outmoded, and the loss of what we believed to be true, what we knew and understood. Not only did it question our goals and the way to reach them, but also it questioned our representation of the world. All this is made more acute by the fourth industrial revolution.

Many studies have been conducted to analyse how this new revolution is likely to impact employment, and even social cohesion, in so-called developed countries. The most anxiety-inducing of these predict a catastrophic outcome: certain professions disappearing en mass, countries destabilized, on the brink of chaos, man being replaced by machines. The more optimistic, however, suggest the transformation of certain ways of working, as has been repeated time and again since Gutenberg, leading to an "augmented" society offering new possibilities.

Divided, quoting contradictory figures, these studies do little to improve clarity. On the more pessimistic side, the very serious Oxford Martin School announces that automation will eliminate one out of two jobs in America during the next twenty years [34]. The no less serious Cabinet McKinsey & Company estimates that 45% of tasks could be automated purely on the basis of present-day experimented technologies, but that only 5% of today's jobs would be at risk [35]: they argue that those jobs eliminated by automation will be thereafter reoriented towards activities with higher added value. The World Economic Forum, on the other hand, predicts that the fifteen most industrialised nations today (with the exception of China) will lose more than 5 million jobs between 2015 and 2020, merely through the impact of new technologies [36]. They also report, as others before them, that two-thirds of the children starting primary school today will end up doing a job that does not yet exist. Along the same lines, the champions of the digital economy consider that in 2030 or thereabouts, 85% of jobs will be totally new and at present totally unpredictable in nature.

The figures and points of view may diverge, but at least they all serve to underscore the present importance of uncertainty.

In parallel, other studies analyse the impact of this new era on the subject of our particular interest: management.

In 2016, the World Economic Forum made a prospective analysis of the year 2020. At such a short horizon their findings are far from divinatory but their report brings a certain clarity to the debate [36]. It does not highlight the emergence of new managerial skills, or the disappearance of prior skills, but rather a redistribution in order of importance, with the resolution of complex problems coming top of the list, followed by critical thinking, creativity and collective and emotional intelligence. Reassuring, politically correct, this conclusion is also very ambiguous: it may encourage us to continue on the basis of our present experience, and not to question anything!

Other, more informative, studies have tried to determine in what way the enterprises of the third industrial revolution are different from those having preceded them [37]. With regard to management, they identify several paradigm evolutions (Table 2):

**Table 2: Changes in paradigm from the 2nd to the 3rd Industrial Revolution**

| EVOLUTION | COMMENTS |
|---|---|
| Guide → Pilot | Do not fixate on a destination but pursue a journey |
| Rigidify → Interact | Do not become overly procedural |
| Assert → Learn | Leave room for doubt, then try to remove it |
| Plan Theorise → Experiment | Shake off intellectual save havens, embrace reality |
| Perfect → Evolve | Do not seek perfection, but adaptation |
| Complicate → Simplify | Embrace complexity, but do not make unnecessary complications |

There is an echo here of the Agile manifesto and its fours cardinal values:

- Individuals and their interactions over processes and tools
- Adapting to change over following a plan
- Customer collaboration over contract negotiation
- A working product over comprehensive documentation

These new paradigms considerably undermine the very image of management; the image managers have of their own role, and also the image as seen from the managees. Over time, generations of managers have been conditioned to reject doubt as being a sign of incompetence: they were supposed to be the source of solutions, results, certainly not doubts – or else, in secret. Here, on the contrary, doubt is expressly introduced into the scope of management: a good manager is one able to embrace the doubt, remove or arbitrate it. Similarly, as seen above, generations of managers have been conditioned to manage risks, via largely illusory approaches. Now, the question is to make the culture of risks evolve towards one of opportunity.

As always, faced with a changing situation, two attitudes tend to be adopted. The first feeds on regret, nostalgia, finds refuge in certainties, beliefs, erstwhile indicators, be they obsolete or illusory, and suffers through the changes. This explains the re-emergence of fundamentalisms and conservatisms of all natures from the onset of the 21st century: when the future seems so obscure, it may seem tempting to seek the light in the tenets, however strict, that had proved their worth in the past. The second attitude, on the contrary, is accepting of the new situation, seeing in it fertile opportunities, stimulating challenges, the pleasure of discovery, and the possibility of deciding one's own destiny. But if this way is chosen, how should we proceed in practice? Must everything be reinvented? Not at all, since the enterprise has long developed means to understand uncertainty and complexity via an activity of which it is the vocation: Research and Development. This represents a major source of precious inspiration to meet the managerial challenges posed by the fourth industrial revolution and, more globally, by the VUCA world.

Curiously, although literature is abundant regarding the new stakes of management, there is little to be found on the contribution of R&D to the debate. Several reasons may explain this anomaly:

- Among the major activities of the enterprise, R&D is often isolated, more secret than the others, saddled with an image both comical and out-dated, that of the stereotypical scatter-brained Professor Nimbus, bubbling retorts and other oddities,

- Among the major processes tackled by re-engineering or by the Lean approach, focus has been put on production, the supply chain, quality of financial control, but not on R&D where activities are less recurrent, its course less formalized and its expected gains less quantifiable,

- Contemporary sensitivity privileges the warmth of human relations, emotions, protection, safety, whereas R&D suffers an image that is radically opposite: cold reason, now unpopular, the unknown, threatening in its very essence, and even danger.

Lastly, and perhaps most importantly, this absence is largely due to drifts in governance. There was a time when, in large companies, the Management Committee had a good operational culture, where the Technical Manager and Industrial Manager wielded the most influence. Then the pressures of marketing and finance shifted the centre of gravity of governance towards other profiles, often divorced from the realities in the field, with little knowledge of the scientific notions inherent to R&D, little inclined to take on board the long lead times of R&D, preferably short-term horizons. In other words, if R&D is part of the DNA of certain champions, it is in many cases marginalized in company culture, and even in action. As regards the companies of the tertiary sector, they have long sailed familiar seas without the need for new sails, and have therefore not often developed any R&D activities worthy of the name.

Thus, R&D seems to be out on a limb, rarely taken as an example in an enterprise. But on the contrary, it could shed useful light on good managerial practices, so necessary in today's world and even more so in tomorrow's. How could this be exploited, out of R&D's particular scope of activity? This is what we will now analyse.

## 4   R&D: INSPIRATIONS FOR CONTEMPORARY MANAGEMENT

We will commence by defining the lexical elements used. Afterwards, we will highlight the similarities between the new forms of entrepreneurship and R&D, and outline the fundamental principles of the R&D process. Finally, we will analyse how these may act as benchmarks for contemporary management and will propose an approach able to lead a strategy to maturity.

## 4.1 Vocabulary

The Frascati manual [38] defines what constitutes R&D, and distinguishes three degrees:

- Fundamental research
- Applied research
- Experimental development

Fundamental research is public by nature, whereas applied research and experimental development concern both the private and public sectors.

The Frascati manual emphasizes uncertainty: to be eligible for the tax advantages offered to R&D, the outcome of a project must not be certain, and its results must stem from the removal of certain technological barriers.

We will deliberately position ourselves in the private sector, which represents 60% of the R&D activity in France and to which the companies of the third industrial revolution belong. In France, the most heavily-investing sectors are respectively: IT (12%), the automobile industry (12%), aeronautics (8%), instrumentation (7%), pharmaceuticals (6%) and electronics (6%).

As far as possible, we will consider Research and Development in all its aspects, but will make the distinction between the different forms of R&D if and where necessary.

Researchers and engineers form the "hard core" of these activities, which correspond to a classification established by the INSEE (French National Institute for Statistics and Economic Research). In addition, R&D draws upon other actors, namely executives, managers and technicians. We will make a point of differentiating between the different roles and responsibilities of these actors.

## 4.2 Similarities between entrepreneurship and R&D

Creating a company and leading an R&D activity share one glaringly obvious element, stepping out into the unknown, but these also share other similarities. At the turn of this century, after having observed the creation of a thousand or so companies in the new economy, S. D. Sarasvathi was able to draw startling conclusions about these successful young companies, which were presented as a principle: the principle of effectuation [39][40]. This is not a theory but an observation based on the intelligent use of feedback. It encompasses the following ideas:

- These companies started out in a small way, by trial and error, and learned from their experience
- They never invested more than they could afford to lose
- They practiced the principle of frugality, later developed by Rajou [41]

We are here a long way from the archetypal practices of the second industrial revolution. Working on the basis of the above ideas, Sarasvathi developed a set of guidelines, dubbed "the pillars of effectuation".

One of these pillars, named the "crazy quilt" consists in multiplying the generation of opportunities. By nature, R&D is a trial-and-error cultural process, the aim of which is to acquire knowledge in an iterative open manner. At each step, a question whose answer is unknown is confronted with an experiment and observations. In most cases, the results of the experiment with which the researcher or engineer validates or invalidates his or her hypotheses generate new knowledge but also new questions, which had not been thought of before: the domain to be explored expands as do the number of opportunities to be seized. In this, the R&D approach is rooted in the "crazy quilt".

Further analysis of the other pillars reveals other points of convergence between entrepreneurship and R&D:

- Bird-in-hand principle: entrepreneurs start with their means, who they are, what they know; R&D, via a study of prior art, does the same before launching any new study,

- Lemonade principle: entrepreneurs consider surprises as a source of opportunity; similarly, R&D embraces the unexpected results of an experiment as an opportunity to discover another aspect of the subject,

- Crazy quilt principle: entrepreneurs build partnerships in the form of a network with varied stakeholders; R&D proceeds in the same way, collaboration between research teams has become a norm,

- Pilot-in-the-plane principle: entrepreneurs focus on the actions and decisions over which they have control and consider that the future cannot be predicted but can be created; in R&D, researchers and engineers know that nature is infinitely complex and that thanks to active experimentation, parts of this complexity can become accessible,

- Affordable loss principle: whatever the project, entrepreneurs ensure that its cost will not jeopardize the viability, namely financial, of the enterprise; R&D is comparably conscious of the necessity for frugality in that experiments are performed on reduced quantities or at a small scale, often by cannibalizing instruments having been used for previous experiments.

Modern entrepreneurship and R&D thus share essential characteristics: the means (what do I know?) define the objectives (what am I doing?) which enable the stakeholders to be identified, who then produce new objectives and inject new means, and hence the cycle continues in successive iterations. Paradigmatically, entrepreneurship by effectuation and R&D are both in opposition to the classical vision of the second industrial revolution, and to the relations between strategy and implementation which are traditionally formed.

- It is the means and the results which define the objective
- It is knowledge which defines the path to be taken over time: there is no point in developing a largely-imaginary plan for the next five years

- It is the opportunities generated by the stakeholders which define the product: the problem is not to imagine a product, not to perfect its design and manufacture it, and then try to find a market for it.

The connection to reality constitutes another field of comparison. In essence, R&D consists in exploring reality: conducting experiments, learning from them, at the price of doubt. In theory, this leaves no place for belief; there is no question of believing but investigating, justifying, interpreting. In practice, this quest is often diverted from its path, researchers and engineers don't see beyond their theories, their conjectures, false certainties or battles of ego, and forget to be objective, impartial and lucid. Similarly, the lucidity of entrepreneurs is permanently tested, leaving them more vulnerable to taking the easy option: rejecting doubt and questioning, the arbitration between several options, uncomfortable and time-consuming, reliance on beliefs and pseudo-certainties presented as truths, even at the price of distorting that truth and or having a blinkered vision of it. The propensity of a company to ignore reality and to follow the safe bet is in itself a mark of its fragility, and may even sign its death warrant. Thus, in the case of KODAK, which we have previously evoked, the company had actually developed its capacity in digital technology, but its governance was locked in in a prison based on two beliefs: that customers would continue to want printed photographs, and that the company could continue to bank on the quality of the silver photography, which had ensured its success in the past, rather than take up the advantages offered by digital photography, namely simplicity (cf. article in "Le Monde" dated 18/11/17). These erroneous beliefs sealed its fate and so its bankruptcy.

Hence, entrepreneurship and R&D share the fact of centring the debate on the connection with reality, of making lucidity one of the most important managerial qualities.

At this point, we will now outline the fundamentals of what we call the "R&D process", so as to extract a source of benchmarking with respect to management as a whole.

## 4.3 Fundamentals principles of the R&D process

To describe the R&D process, we will deliberately avoid using a linear approach, which would be contrary to the modalities of R&D that by nature are iterative. Instead, we will adopt an approach according to the following themes.

### Strategy

An R&D activity may have very diverse perimeters and magnitudes, from the study of a technology brick to designing a complete system. It may originate from one of several sources: an expression of requirements from programme management, the anticipation of a future need from sales management or from a prospective team, or else a "blueprint" for technological development emitted from the R&D management itself. It may be financed by a contractor, be self-financed or be a combination of both.

### Organisation

Generally speaking, a distinction must be made between the different activities of R&D. If they concern Research, the activities are often conducted within largely autonomous technical or R&D departments. If they concern Development, their multidisciplinary nature usually requires different professional areas and skill sets to be mobilized and thus also the resources belonging to different sectors. In this case, a transversal organisation must be set up, and two main models are used: the project mode and the matrix organisation.

Research in the public domain has not made the same organizational choices and remains, in the majority, organized along the lines of 19$^{th}$ sciences, after which the different laboratories are named. In the private sector, innovation managements or business units group functions that go way beyond scientific disciplines to include legal, commercial and support functions. Situated upstream, they are careful to coordinate their activities with those downstream, namely industrialization and marketing.

Each mode of organization suffers its own shortcomings: an organization based on autonomous lines of work can promote corporatism, a project organization runs the risk of generating a heterogeneous approach from one project to another, thus hindering the sharing of knowledge, a matrix organization always raises the problem of the availability of resources, torn between their contribution to the project and their sector of origin.

Organizing the governance if a key issue, and this from the onset. When the activity is self-financed Research, the governance often remains within the entity concerned. If there is an internal specifier, for example programme or innovation Directorate, the governance is widened to integrate it. When the activity deals with Development, the governance is further widened to include the various internal stakeholders. In any event, certain choices are crucial: a clear "project holder" must be defined, decision-makers designated, means allocated to operational teams, and steering and reporting modalities determined. In the case of project or matrix organizations, a leader must be designated who not only enjoys full legitimacy but, beyond being officially designated, genuine leverage capability.

### Operations

Any R&D activity requires planning. Indeed, as already seen, R&D is a journey built of questions, proofs, firm answers, partial answers and non-answers, of successes and failures, certainties and doubts, generating other questions. Prerequisite to any scientific approach, this journey must inevitably be structured: it is mandatory to know where we are with respect to others (state of play vs state of art), to organize the questioning, analyses, trials and proofs, to define the different steps which gradually allow gaining in knowledge and maturity. For this last purpose, R&D often uses the Technology Readiness Level or TRL scale [42]. This comprises 9 levels extending from a simple basic premise to proven technological mastery enabling mass production and commercialization. Contrary to planning approaches already mentioned, and namely that of strategic planning or risk engineering, this planning approach is strictly operational and reality-based. In practical terms, it leads to drafting a reference document that describes the intellectual approach to be followed, the way along which the different steps will fit together, the decision milestones, the type of trials, the associated means, the way in which the trials and their results will be documented, the deliverables to be produced. This document may be of various forms depending on the nature of the activity, and may include anything from an experiment plan comprising a handful of pages to a full development plan. The approach also integrates quality criteria, essential to activities where bugs are unacceptable and where error can be fatal.

Such focus on planning may be a source of paradoxes and, sometimes, of misunderstanding: R&D is also a matter of gushing, being intuitive, and tackling what is unforeseen! The art of R&D actors consists in taking a direction but being able to change it whilst keeping the target in mind, in being able to combine opposite qualities, methodology and flexibility, rigour and pragmatism, action and analysis.

### Steering

Ensuring a proper steering of the activity or project is a further key issue. This means ensuring that the decisions are made at the right time, with sufficient inputs. Planning thus integrates milestones where decisions may be made and supported, according to the maturity of the activity or project.

### Methodology

Finally, it is fairly obvious that R&D is based on a set of methodologies. Without claiming to be exhaustive, three points seem to be particularly important:

- A community between actors needs to be put in place to foster emulation and encourage ideas to circulate. This involves regular but flexible get-togethers, discussions where diverse viewpoints could be freely confronted, outside of "institutionalized" meetings. It also involves the use of certain tools, shared databases, an exchange platform unique to the community and not necessarily integrated in the company information system, but nevertheless having a properly managed "legal" existence.

- The context of the experiments needs to be fully documented, without which no sound conclusion could be drawn. This means, in particular, extensively documenting the trial protocols, the configuration of materials tested, and the means implemented.

- Experiment feedback should be formalised, shared and capitalized. The objective is to draw conclusions, ensure they are explained, that they do not remain known to only a few, that they are accessible in the long-term. This forms part of knowledge management and requires a structured knowledge base.

## 4.4 Benchmark with respect to management

Essentially, R&D lies at the heart of our representations of the world and questions such representations in a structured manner that combines thought and action. This structure may be divided along three main lines (Figure 2):

- Cultural processes
- Operational processes
- Governance (decision-making)

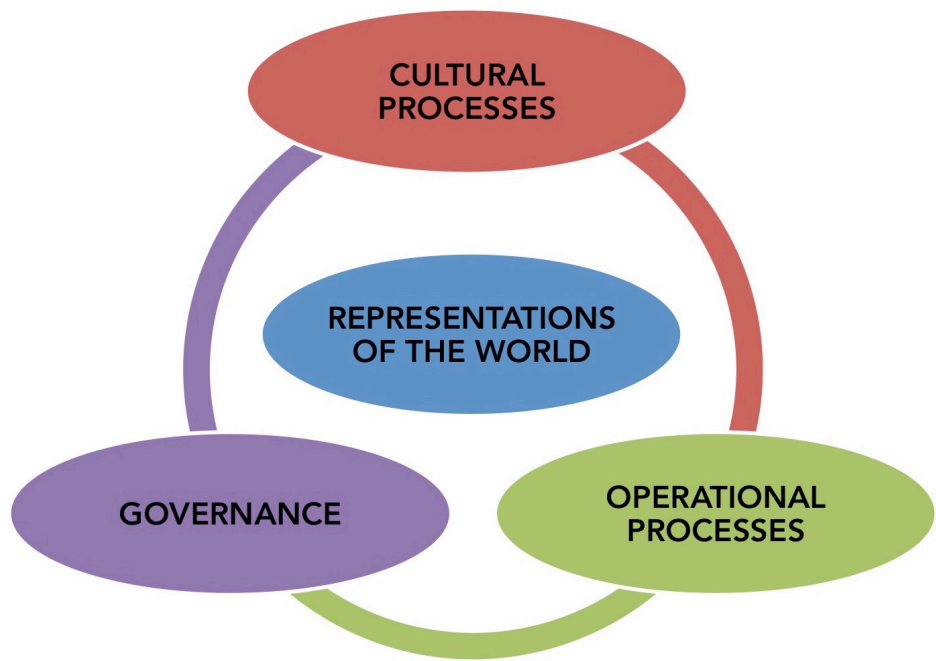

**Figure 2: main lines of R&D**

From this we can draw the following sources of inspiration:

### Cultural Processes

1. Leverage questioning as well as finding, learning as well as expertise, knowledge as well as solutions
2. Extend the notion of performance to integrate that of lucidity, creativity, cooperation, aid in decision-making
3. Leverage progress rather than perfection
4. Invest in investigation and experimentation rather than fall back on habits, theories, following where others have gone before
5. Invest in feedback management (for successes as well as failures): formalise, capitalise, share
6. Encourage critical thinking, diversity, positive contradiction and reject one-track thinking, uniformity, conformity

7. Recognise the right to doubt, including of oneself
8. Recognise the right to be wrong, provided that lessons are learned

**Operational Processes**

1. Stimulate cooperation and cross-thinking by means of shared projects
2. Provide both a framework and room for initiative
3. Structure the exploratory approach and adjust it according to circumstance
4. Characterise the context: the environment, the test conditions, the potentially influential parameters, etc.
5. Cultivate objectivity: facts, results, interpretation
6. Assess progress using factual criteria based on the TRL scale
7. Establish shared tools

**Governance**

1. Develop a global vision (systemic approach)
2. Try to develop lucidity and foresight rather than reassurance
3. Rely on skills, not institutions
4. Rely on proofs, not impressions
5. Organise decision milestones corresponding to events, not to an administrative calendar
6. Give the operational actors a role in decision-making, rather than restricting them to merely following orders

In all this, the cultural dimension is paramount. Depending on whether the enterprise is capable of tackling reality or falls back on the reassurance of illusion, it will head towards foresight or towards approximation, dead-ends, and errors. Depending on whether it has the courage to face the long-term or falls back on the short-term, admits or refuses doubt, failure, it will not face the same choices. Whether it allows bottom-up mode or works strictly top-down, accepts or rejects criticism, encourages diversity or maintains uniformity, it will enlarge the realm of possibilities or will shrink it. R&D thus embodies a double inversion with respect to traditional management modalities:

- Cultural dimensions take precedence over operational processes, even if these are quite naturally crucial (the methodology must be professional)
- Governance is no longer the keeper of knowledge, this is held by the operational actors. Governance finds its power and legitimacy in its ability for to provide impetus, listen, analyse, arbitrate, widen options, change course and provide support

This inversion may be displayed in the following way (Figure 3).

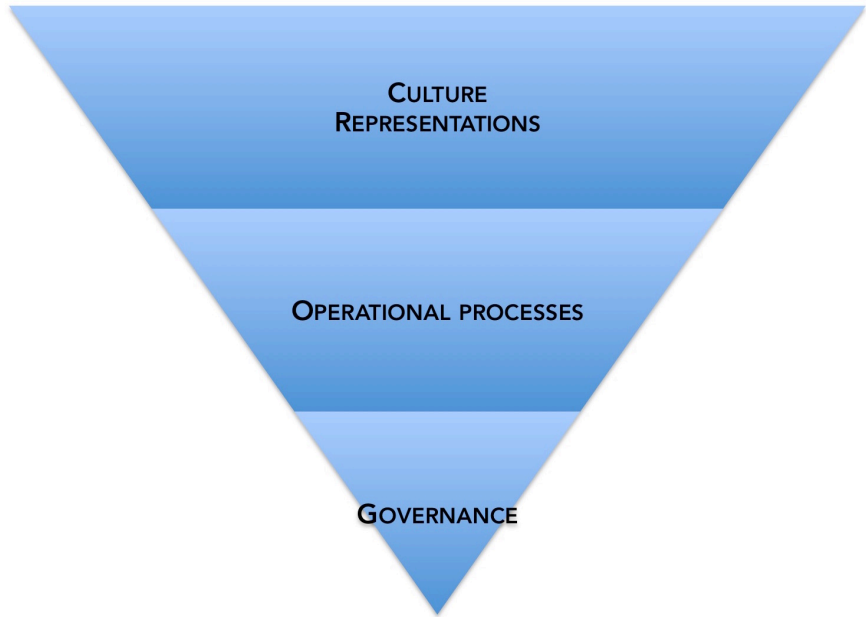

Figure 3: Inversion of management modalities

All of this stems from representations of the world, fosters them, makes them evolve, and brings the enterprise at strategic level.

**Strategy**

There again R&D embodies an inversion with respect to traditional management modalities: it is no longer a question of establishing strategy in a closed loop, then communicating it from top to bottom, but of building strategy as part of an ongoing exploratory process, enabling it to steadily mature and confront reality. By extension of the TRL scale of technological maturity used in R&D, we propose to create a scale of maturity with respect to strategy, the Strategy Readiness Level or SRL:

- SRL I: Opportunity detection and analysis
    - An unproven opportunity is detected (solvency unknown, feasibility unknown): it is the origin of an *intent*
    - On paper, the means by which a solvent customer requirement could be fulfilled is imagined
    - A first Osterwalder business model is produced but without completing all the fields: a search for the missing information is launched
- SRL II: Inflation of knowledge
    - The different fields of the business model have been completed, but have not yet been organised or checked for consistency
    - The experiments required to raise business uncertainties or remove technological barriers are defined and their cost calculated
- SRL III: Experiments and first contact with the market
    - The experiments are performed on a small scale and completed by a creativity approach (Design thinking [43], User experience [44]…); these

put the customer in the loop who thus becomes a major stakeholder
        according to Porter's forces
    - The first market tests are performed
- SRL IV: Design and industrialization
    - The product undergoes more detailed preliminary designing
    - Its industrialization is integrated as early as possible into the design process
    - At the key stages in the development plan, release candidate products are launched onto the market, their configuration and the design justifications are fully documented: related feedback is taken into consideration for the following iteration
    - Lastly, when deemed mature enough, the product begins its commercial life

The originality of this new approach lies in the continual coupling of strategy and implementation, rather than their traditional decoupling.

# 5 CONCLUSION

The second industrial revolution, followed by the digital revolution of the past thirty years, has led to ever-growing complexity and uncertainty, in parallel with the evolution of society as a whole, which is itself also marked by its complexity and uncertainty. The contemporary world, the on-going revolution and the arrival of artificial intelligences raise the emergent question of complexity and uncertainty to unprecedented heights.

Managerial evolutions since the beginning of the 20th century have been legion, characterized on the "hard" side by the end of the Taylorist approach, the rise of systemic management, strategic management, and performance management, and on the "soft" side by the entry in force of the human dimensions in business, sociology, psychology, human relations, motivations, and emotions. The hard-based evolutions have revealed their limitations: the segregation between the governance and operational actors at working levels, an inflation of rules, procedures and inspections, pressure from the short-term and the relentless pursuit of efficiency, all leading to a loss of meaning, responsibility, pleasure, commitment, and the record development of psycho-social risks. Human-based evolutions, like the quality of work life, the liberated or "opal" company, are currently on a roll, but often have the disadvantage of opposing the human dimension and the operational dimension, discredited by errors made on the "hard" side. A first major issue for contemporary management, we think, consists in reconciling these two dimensions, "soft" and "hard".

Beyond that, complexity and uncertainty seriously undermine our representations of the world. What do we really know? What could happen next? On what can we ground our decisions? Overwhelmed by these uncomfortable questions, we are tempted to reminisce about the world as it was and try to seek refuge in past reassurances, established beliefs, familiar conventions, even false certainties and illusions. There is another path: that of accepting these mutations, restoring the virtue of doubt, of critical thinking, questioning and harnessing them to reach a true *awareness*. This is a second major issue for contemporary management.

Faced with these two issues, it is more than ever necessary to stop following the latest trends, stop following in others' footsteps: this is by nature incompatible with complexity and uncertainty. Instead, it would be preferable to look for sources of *inspiration*. Research and Development offers a particularly pertinent source of inspiration, since its very purpose is to understand the complex and uncertain, to explore doubt, to support viewpoints, to gain knowledge and to learn lessons from both success and failure.

Its core principles decoded, R&D can act as a benchmark for contemporary management:

- It embodies a profound evolution in managerial paradigms: no longer fix a destination but prepare a *journey*, no longer plan or theorize but experiment, no longer assert but learn, no longer set but adjust, no longer act but interact.
- It solves the tension between the "hard" and "soft" aspects by positioning the cultural dimension at the top of the hierarchy; not in the form of oft-vaunted generic values but in terms of operational values which emphasize questioning, learning, experimentation, skills, the sharing of knowledge, the diversity of viewpoints, the freedom to disagree.
- It rehabilitates management by processes, so often disparaged, by making a distinction between processes (what?) and procedures (how?) and by drawing on best practices.
- It embodies a different logic curing the governance drifts that have been widely observed, where it is no longer a question of deciding "at the top" and following order "at the bottom" (top-down) but considering operational actors as aids in decision-making and associating them (bottom-up).

What could the advantage of exploring, of better understanding? How do we begin? How can we develop true viewpoints? How can we make decisions without knowing all the facts? The R&D approach addresses these crucial questions, not only from an R&D perspective but to the advantage of the enterprise as a whole.

If innovation is not the sole province of technology, it is also true that the absence of technology in a project makes it vulnerable to new entrants [32]. If a project or product is not linked to a unique know-how, protected by patent and/or kept in secret, then any other investor may copy it. This ability to eliminate entry barriers as described by Porter constitutes in itself a justification for R&D.

Seeing in R&D a source of inspiration, however, must not blinker us to the common trap of transforming it into a model and blindly reproducing it. The approach consists firstly in building a culture of questioning, doubt and lucidity, in contradiction to prevailing management values: assertion, certainty, the illusion of risk management, dashboards truncating or masking reality. Swimming against this particular tide requires a strong will.

Paradoxically, it is important not to trivialise what it represents. Indeed, the term "innovation" has become a watchword of communication, ranking high in many websites, sales brochures or plenary speeches, but its catch-all use has diluted the strength of its meaning. The term R&D should therefore not become just another buzzword, but embody a true managerial approach.

Lastly, if R&D offers the advantage of supplying a structured approach to question our representations of the world, it is not the sole discipline to do so: this is also the vocation of sciences in general, philosophy, spirituality, and politics in the true sense of the term. Ultimately, all these paths commonly question power, control, how to reach a proper balance between the goal and the journey, between intent and hesitancy, between will and abandon.

**References**

[1]     F. W. Taylor, *The principles of scientific management*. Sittard: Management Classics, 2015.

[2]     H. Fayol, *Administration industrielle et générale - Prévoyance, organisation, commandement, coordination, contrôle*. H. Dunod et E. Pinat, 1917.

[3]     M. Weber and M. Weber, *Les catégories de la sociologie*. Paris: Pocket, 2008.

[4]     M. P. Follett, *Creative experience*. Mansfield Centre, CT: Martino Publishing, 2013.

[5]     D. McGregor and C. MacGregor, *The professional manager*. New York: McGraw-Hill, 1967.

[6]     P. F. Drucker, *The practice of management*. New York, NY: HarperCollins, 2009.

[7]     M. Crozier and E. Friedberg, *The bureaucratic phenomenon*. New Brunswick: Transaction Publ, 2010.

[8]     M. Crozier and E. Friedberg, *L'acteur et le système les contraintes de l'action collective*. Paris: Éd. du Seuil, 2014.

[9]     M. D. Cohen, J. G. March, and J. P. Olsen, "A Garbage Can Model of Organizational Choice," *Adm. Sci. Q.*, vol. 17, no. 1, p. 1, Mar. 1972.

[10]    J.-C. Fauvet, *L'élan sociodynamique*. Paris: Éditions d'Organisation, 2004.

[11]    H. Mintzberg, *Grandeur et décadence de la planification stratégique*. Paris: Dunod, 2004.

[12]    "State of the Global Workplace - Gallup 2013." .

[13]    B. M. Carney, I. Getz, and O. Demange, *Liberté & cie: quand la liberté des salariés fait le succès des entreprises*. S.l.: Flammarion, 2016.

[14]    F. Laloux, *Reinventing Organizations: Vers des communautés de travail inspirées*. 2015.

[15]    R. K. Greenleaf and L. C. Spears, *Servant leadership: a journey into the nature of legitimate power and greatness*, 25th anniversary ed. New York: Paulist Press, 2002.

[16]     G. Verdier et N. Bourgeois, *Libérer l'entreprise ? Confiance, responsabilité et autonomie au travail*. Paris : Dunod, 2016.

[17]     K. Schwab, *The fourth industrial revolution*, First U.S. edition. New York: Crown Business, 2017.

[18]     L. Ferry, *La révolution transhumaniste: comment la technomédecine et l'uberisation du monde vont bouleverser nos vies*. Paris: Plon, 2016.

[19]     N. Bennett and G. J. Lemoine, "What a difference a word makes: Understanding threats to performance in a VUCA world," *Bus. Horiz.*, vol. 57, no. 3, pp. 311–317, May 2014.

[20]     P. Silberzahn, "Blog de P.silberzahn https://philippesilberzahn.com." .

[21]     G. Nicolis, I. Prigogine, and P. Carruthers, "Exploring Complexity: An Introduction," *Phys. Today*, vol. 43, p. 96+, 1990.

[22]     H. C. Lucas and J. M. Goh, "Disruptive technology: How Kodak missed the digital photography revolution," *J. Strateg. Inf. Syst.*, vol. 18, no. 1, pp. 46–55, Mar. 2009.

[23]     L. Brunet, J.-P. Guillemin, and G. Hayotte, "La métrologie de l'anticipation : capter les risques émergents," in *6eme colloque capteur*, 2008.

[24]     M. Halévy, *Un univers complexe: l'autre regard sur le monde*. Escalquens: Oxus, 2011.

[25]     F. H. Knight, *Risk, uncertainty and profit*. Kissimmee, Fla: Signalman, 2009.

[26]     A. Osterwalder, "The business model ontology a proposition in a design science approach," Ecole des Hautes Etudes Commerciales de l'Université de Lausanne, Lausanne, 2004.

[27]     P. Thiel and B. Masters, *Zero to one: notes on startups, or how to build the future*. New York: Crown Business, 2014.

[28]     S. G. Isaksen and Gaulin, "A Reexamination of Brainstorming Research."

[29]     B. Mullen, C. Johnson, and E. Salas, "Productivity Loss in Brainstorming Groups: A Meta-Analytic Integration," *Basic Appl. Soc. Psychol.*, vol. 12, no. 1, pp. 3–23, Mar. 1991.

[30]     L. E. Brunet, "Créativité en ingénierie," in *ref. article : ag5210*, vol. base documentaire : TIB127DUO, T. de l'ingénieur S. de conception pour l'innovation, Ed. 2015.

[31]     J. Kruger and D. Dunning, "Unskilled and unaware of it: How difficulties in recognizing one's own incompetence lead to inflated self-assessments.," *J. Pers. Soc. Psychol.*, vol. 77, no. 6, pp. 1121–1134, 1999.

[32]     L. E. Brunet and K. Le Meur, "A big data and Darwinian approach of scientific creativity," in *R&D Management Conference proceedings*, Stuttgart, 2014, vol. 1, pp. 401–407.

[33]     M. E. Porter, "How competitive forces shape strategy," 1979.

[34]     C. B. Frey and M. A. Osborne, "The future of employment: How susceptible are jobs to

computerisation?," *Technol. Forecast. Soc. Change*, vol. 114, pp. 254–280, Jan. 2017.

[35]     "Future that works : automation, employment and productivity." Mc Kinsey, 2017.

[36]     "The future of jobs : Employment, Skills and Workforce Strategy for the Fourth Industrial Revolution." 2016.

[37]     G. Dosi and L. Galambos, Eds., *The third industrial revolution in global business*. Cambridge: Cambridge University Press, 2013.

[38]     "Frascati Manual: Proposed Standard Practice for Surveys on Research an...," May-2015. .

[39]     S. D. Sarasvathy, "Causation and Effectuation: Toward a Theoretical Shift from Economic Inevitability to Entrepreneurial Contingency," *Acad. Manage. Rev.*, vol. 26, no. 2, pp. 243–263, Apr. 2001.

[40]     S. D. Sarasvathy, "What makes entrepreneurs entrepreneurial?," 2001.

[41]     N. Radjou, J. Prabhu, and S. Ahuja, *Jugaad innovation: Think frugal, be flexible, generate breakthrough growth*. John Wiley & Sons, 2012.

[42]     "ISO/FDIS 16290 - Space systems – Definition of the Technology Readiness Levels (TRLs) and their criteria of assessment."

[43]     *Design Thinking: Understand Improve Apply (Understanding Innovation)*, 1st Edition. Springer, 2010.

[44]     M. Obrist, V. Roto, and K. V. V. Mattila, "User experience evaluation: do you know which method to use?," in *Proceedings of the 27th international conference extended abstracts on Human factors in computing systems*, Boston, MA, USA, 2009, pp. 2763–2766.